\documentclass[10pt,aps,prb,twocolumn]{revtex4-1}
\usepackage{graphicx}
\usepackage{amssymb}
\usepackage{amsmath}
\usepackage{appendix}
\usepackage{comment}

\def\p{\partial}

\def\bm{\mathbf{m}}
\def\br{\mathbf{r}}
\def\bh{\mathbf{h}}
\def\bj{\mathbf{j}}
\def\bk{\mathbf{k}}
\def\bn{\mathbf{n}}

\def\bp{\mathbf{p}}
\def\bs{\mathbf{s}}

\def\bz{\mathbf{z}}
\def\bx{\mathbf{x}}
\def\by{\mathbf{y}}
\def\bA{\mathbf{A}}
\def\bE{\mathbf{E}}

\def\bl{\mathbf{l}}
\def\bL{\mathbf{L}}
\def\bM{\mathbf{M}}
\def\bH{\mathbf{H}}
\def\bB{\mathbf{B}}

\def\bS{\mathbf{S}}

\def\BN{\boldsymbol{\nabla}}
\def\BE{\boldsymbol{\varepsilon}}

\def\BS{\hat{\boldsymbol{\sigma}}}
\def\BT{\hat{\boldsymbol{\tau}}}

\def\mF{\mathcal{F}}

\def\mK{\mathcal{K}}

\def\mM{\mathcal{M}}

\def\mP{\mathcal{P}}

\def\mT{\mathcal{T}}
\def\mU{\mathcal{U}}

\newcommand{\rf}[1]{(\ref{#1})}
\newcommand{\an}[1]{\begin{align}#1\end{align}}
\newcommand{\al}[1]{\begin{aligned}#1\end{aligned}}
\newcommand{\eq}[1]{\begin{equation}#1\end{equation}}
\newcommand{\seq}[1]{\begin{subequations}#1\end{subequations}}

\newcommand{\ket}[1]{|{#1}\rangle}
\newcommand{\av}[1]{\langle{#1}\rangle}

\usepackage{mdframed}

\begin{document}

\title{Magnetic dynamics with Weyl fermions}

\author{Yaroslav Tserkovnyak}
\affiliation{Department of Physics and Astronomy, University of California, Los Angeles, California 90095, USA}

\begin{abstract}
Transport of charge and valley degrees of freedom coupled to order-parameter dynamics in magnetic Weyl semimetals is studied in the framework of nonequilibrium thermodynamics. In addition to the established valley-related transport anomalies that are rooted in band-structure topology, we construct dissipative couplings between the three dynamic constituents of the system driven out of equilibrium by electromagnetic perturbations. We show how the valley degree of freedom mediates an effective coupling between the charge and magnetic sectors of the system, through a combination of the chiral anomaly, on the electric side, and the Onsager-paired valley torque and pumping, on the magnetic side. This work compliments previous studies of magnetic Weyl semimetals by a more systematic analysis of collective dissipation. We discuss several concrete examples of the valley-mediated current-driven magnetic instabilities and charge pumping, and extend the theory to the antiferromagnetic case.
\end{abstract}

\maketitle

\section{Introduction}

Interplay between collective and incoherent degrees of freedom has been a central theme in spintronics and nonequilibrium magnetism for decades.\cite{wolfSCI01,*zuticRMP04,*sinovaNATM12} An important example of this is provided by itinerant electron transport in natural or engineered metal-based magnetic structures, which can drive magnetic order-parameter dynamics and reciprocally be pumped by the latter, while also experiencing a range of magnetotransport and magnetoelectric phenomena.\cite{tserkovRMP05,*brataasPRP06,*ralphJMMM08} More recently, electronic quasiparticles have been replaced by magnons and other spin excitations, which significantly broadened the scope of the coupled transport and dynamics in magnetic systems towards insulator-based heterostructures.\cite{tserkovJAP18} This was especially fruitful in regard to heat-controlled spin flows and their coupling to collective magnetic dynamics,\cite{bauerNATM12,*kovalevEPL12,*heremansPHYS14} which generally obviates a need for charge carriers.

A natural inquiry then arises concerning other types of transport phenomena and the underlying elementary degrees of freedom in regard to more general controls and detection of order-parameter dynamics. Weyl semimetals provide a fertile ground to this end, as their strong spin-orbit interactions and the emergent valley degree of freedom set the stage for an intricate interplay between charge, valley, and magnetic dynamics (supposing the time reversal is spontaneously broken by, e.g., a ferro- or antiferro-magnetic order).\cite{armitageRMP18,*arakiANP20}

Here, we develop a general phenomenological framework for the coupled dynamics between charge and valley flows, on the one hand, and magnetic dynamics, on the other, in the presence of electromagnetic fields. In addition to the established chiral anomaly associated with Weyl fermions and the anomalous Hall effect in the presence of magnetism,\cite{armitageRMP18,*arakiANP20} general dissipative couplings between the constituent degrees of freedom are constructed, as dictated by structural symmetries and Onsager reciprocity.\cite{landauBOOKv5} At the heart of our key findings are the torques on the order-parameter dynamics by valley transport, along with the reciprocal valley-motive forces. This work illustrates how the itinerant valley degrees of freedom arising in topological semimetals can efficiently replace the conventional spin transport in controlling and responding to collective magnetic dynamics. Our main focus is on ferromagnetic materials, with a generalization to antiferromagnets briefly discussed at the end. Some practical examples addressed in the paper concern valley-mediated current-induced instabilities and switching of the magnetic order, as well as reciprocal valley pumping and its electrical signatures.

The article is organized as follows: In Sec.~\ref{es}, we briefly review a construction of a minimal four-band model for a Weyl semimetal, with spatial inversion symmetry and ferromagnetic order, along with its (anomalous) quasiequilibrium response to electromagnetic fields. Section~\ref{dr} presents our main results concerning the coupled charge-valley-magnetic dynamics, in the form of nonequilibrium thermodynamics.\cite{landauBOOKv5} Section~\ref{ex} is devoted to several specific illustrative examples and, in Sec.~\ref{af}, we discuss a generalization to a dynamic antiferromagnetic order, followed by concluding remarks in Sec.~\ref{sd}.

\section{Electronic structure}
\label{es}

\subsection{Minimal four-band model}

We will develop our phenomenology from the ground up, starting with a minimal microscopic model for a pair of Weyl points in three spatial dimensions. To this end, let us expand a four-band $\bk\cdot\bp$ theory\cite{landauBOOKv9} near the $\Gamma$ point (i.e., $\bk\to0$) of the Brillouin zone. In order to separate the Weyl points, we assume a time-reversal symmetry breaking by a uniform magnetic order parameter $\bM$. Other than that, the system is inversion symmetric and axially symmetric (around the $z$ axis). While this is not essential, we can, furthermore, suppose it is completely isotropic.

Starting with a Bloch state $\ket{\uparrow,+}$ at the $\Gamma$ point, which realizes the usual spin-$1/2$ representation under $z$ rotations $\mU_z$ by angle $\theta$ (which might in general be discrete, as in the cases of three- or four-fold symmetries, for example): $\mU_z(\theta)\ket{\uparrow,+}=e^{-i\theta/2}\ket{\uparrow,+}$, we define the other three basis states as $\ket{\downarrow,+}\equiv\mT\ket{\uparrow,+}$, $\ket{\uparrow,-}\equiv \mP\ket{\uparrow,+}$, and $\ket{\downarrow,-}\equiv\mP\mT\ket{\uparrow,+}$. Here, $\mT$ is the time reversal and $\mP$ is the parity transformation, which obey $\mT^2=-1$,  $\mP^2=1$,  and $[\mP,\mT]=0$. Also, $[\mP,\mU_z(\theta)]=0$ and $[\mT,\mU_z(\theta)]=0$. We suppose these four states, $\ket{s,\tau}$, $s=\uparrow,\downarrow$ being the effective spin and $\tau=\pm$ valley labels, arise in the low-energy long-wavelength description of the electronic structure in the absence of magnetism, where the crystal is symmetric under both $\mP$ and $\mT$. The time-reversal operator, in our basis, acquires the conventional form $\mT=i\hat{\sigma}_y\mK$, where $\mK$ is complex conjugation, while $\mP=\hat{\tau}_x$, using $\BS$ ($\BT$) to denote the vector of Pauli matrices in the spin (valley) space.\cite{Note1}

\begin{figure}[!pt]
\includegraphics[width=0.9\linewidth]{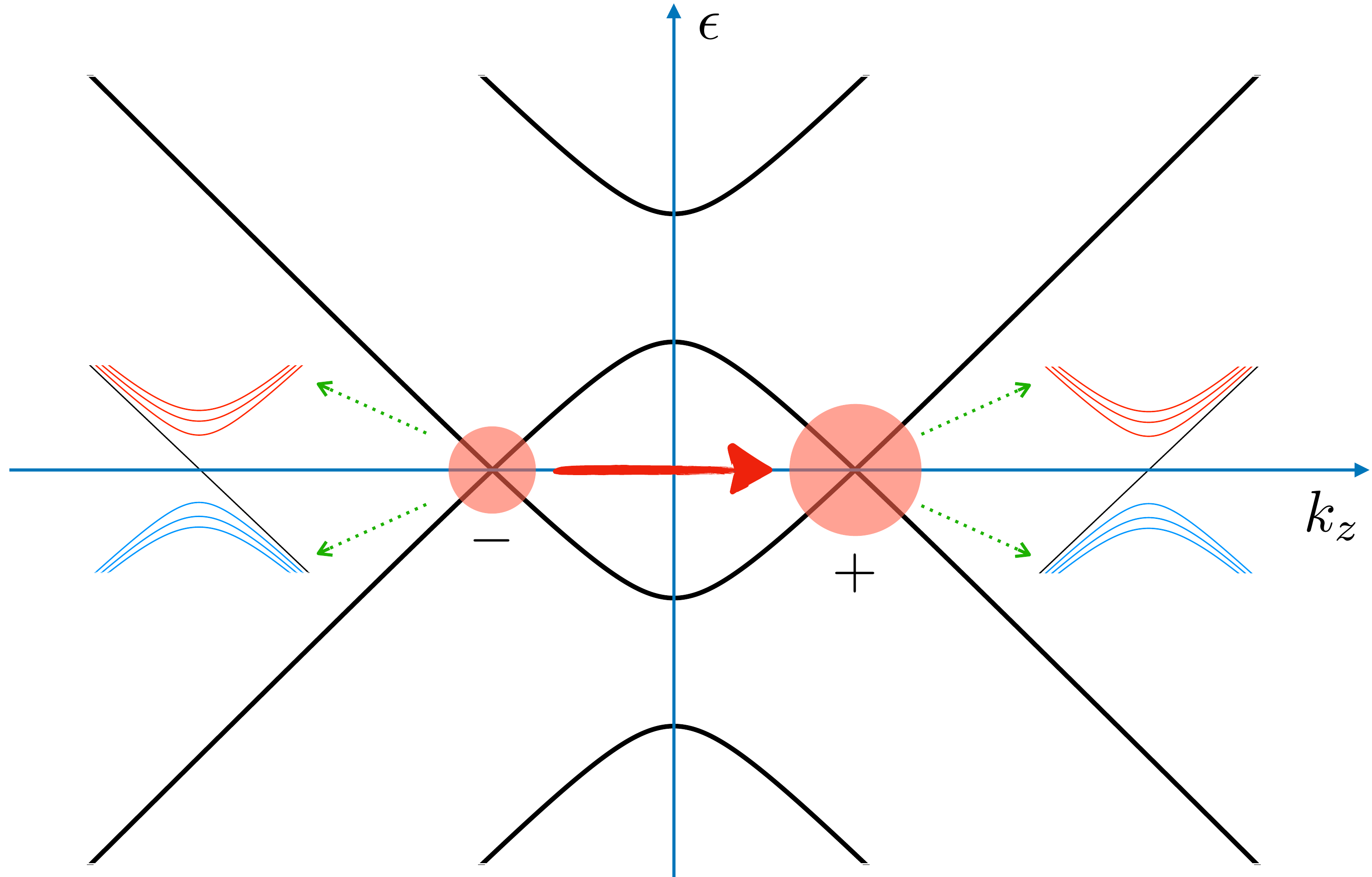}
\caption{The dispersion \rf{ek}, setting $J=3\Delta$, along the $k_z$ axis. The Bloch states at the two, $\pm$ Weyl points, $k_z=\pm\sqrt{J^2-\Delta^2}$ (setting here $v=1$ and $\bM=\bz$), are related by the parity transformation. The solid circles at the two Weyl points schematically designate the corresponding chemical potentials, out of equilibrium. The red arrow depicts the quantized Berry flux through a plane normal to $k_z$ inserted between the two Weyl points (which act as Berry-flux monopoles).\cite{armitageRMP18} The zoom-out insets flanking the Weyl points show the Landau-level dispersions \rf{ll} in the presence of a magnetic field along the $z$ axis (with the monotonic, linear branches stemming from the zeroth Landau levels).}
\label{disp}
\end{figure}

These ingredients allow us to expand the $\bk\cdot\bp$ Hamiltonian to linear order in $\bk$ and $\bM$ as
\eq{
H=v\hat{\tau}_z\BS\cdot\bk-J\BS\cdot\bM+\Delta\hat{\tau}_x\,,
\label{H}}
which provides a minimal model for two Weyl points coupled to magnetic dynamics. One can check all the invoked symmetries, recalling that both $\bk$ and $\bM$ flip under time reversal, while $\bk$ behaves as a vector and $\bM$ as a pseudovector under spatial inversion. $\BT$ and $\BS$ commute, while $\hat{\tau}_z$ flips under inversion: $\hat{\tau}_x\hat{\tau}_z\hat{\tau}_x=-\hat{\tau}_z$. $v$, $J$, and $\Delta$ are symmetry-governed real-valued parameters. For $\bM=\bz$ and absorbing $v$ in $\bk$, one finds the spectrum (by squaring $H$) to be
\eq{
\epsilon_\bk=\pm\sqrt{k^2+\Delta^2+J^2\pm2J\sqrt{k_z^2+\Delta^2}}\,.
\label{ek}}
When $\Delta=J=0$, we get a double-degenerate Dirac cone at $\bk=0$. $\Delta\neq0$ is an ordinary Dirac mass term, such that when $|\Delta|>|J|$, we get a (featureless) gapped insulator. Most interestingly for us, when $|J|>|\Delta|$, the gap closes and we restore a semimetallic band structure, but now with two Weyl points on the $z$ axis at $k_z=\pm\sqrt{J^2-\Delta^2}$ (see Fig.~\ref{disp}).

Once the Weyl points appear, they are topologically robust against small variations in the parameters of the Hamiltonian, irrespective of the underlying symmetries.\cite{armitageRMP18} This sets the stage for our discussion of the collective response of the electronic system to and its feedback on a smooth spatiotemporal magnetic texture. The key physical consequences of the Weyl fermions, such as the chiral magnetic effect and the anomalous Hall effect, are, furthermore, qualitatively insensitive to the microscopic details,\cite{armitageRMP18} so that our particular toy model is invoked mainly for illustrative purposes.

\subsection{Electromagnetic coupling}

Minimally coupling the electromagnetic field to the electrons, the Hamiltonian becomes
\eq{
H(\bk)\to H(\bk-\bA(\br,t))+V(\br,t)-\mu\BS\cdot\bB\,,
}
where the 4-potential $(V,\bA)$ parametrizes the physical electric, $\bE=-\BN V-\p_t\bA$, and magnetic, $\bB=\BN\times\bA$, fields (setting $\hbar=e=c=1$; the sign of the electron charge being absorbed in the electromagnetic fields). The last term in the above Hamiltonian is due to a magnetic moment $\mu$ that may be associated with the effective spin $\BS$. This Zeeman term can be combined with the magnetic-exchange piece $\propto J$ of the Hamiltonian \rf{H}, giving a correction that we will disregard in the following. The exchange term, in turn, can be absorbed into the shift of the vector potential:
\eq{
\bA\to\bA+\frac{J}{v}\hat{\tau}_z\bM\,.
\label{A}}

We suppose, hereafter, that $J|\bM|\gg\Delta$, which makes $\hat{\tau}_z$ approximately conserved and allows us to treat $\Delta$ as a small parity-mixing perturbation. The Weyl-point valleys are now approximately placed at $\bk_\pm=(J/v)\bM$, with the $\Delta$ term allowing to mix them, when complemented with the inevitable momentum-scattering disorder or inelastic scattering. We will treat this as a weak process, endowing the valley index with a finite lifetime $\tau_v$.

Two universal features follow directly from the above band structure. First of all, the quantized Berry curvature emanating from the Berry-flux monopoles at the Weyl points engenders the \textit{anomalous Hall effect.}\cite{nagaosaRMP10} The Berry-curvature flux (for the occupied bands) through a plane between the two Weyl monopoles shown in Fig.~\ref{disp}, with the Fermi level close to zero energy, is fixed at $2\pi$ (i.e., Chern number 1).\cite{armitageRMP18} The integrated Hall conductance $\propto(\bk_+-\bk_-)/2\pi$ results in the Hall (particle) current density
\eq{
\bj^{(H)}_c=\frac{\eta}{2\pi^2}\bm\times\bE\,,
\label{jH}}
in the presence of an electric field $\bE$. Here, we expressed the order parameter as the collective \textit{spin density,} $\bM=s\bm$, in terms of its magnitude $s$ and direction $\bm$ (such that $|\bm|\equiv1$), and defined
\eq{
\eta\equiv\frac{Js}{v}\,.
}

Secondly, the Weyl points are associated with a \textit{chiral anomaly}.\cite{armitageRMP18} Namely, a uniform electronic state, subjected to a DC magnetic field $\bB$, builds up valley imbalance at the rate
\eq{
\p_t\rho_v\to\frac{1}{2\pi^2}\bE\cdot\bB\,,
\label{EB}}
when subjected to a uniform electric field $\bE$. We define the valley density $\rho_v$ as the difference of particle densities assigned to the $\pm$ valleys (relative to the Weyl points): $\rho_v\equiv\rho_+-\rho_-$. Related to this, shifting the electronic (valley) chemical potential $\mu_c$ ($\mu_v$) results in the valley (particle) current density
\eq{
\bj^{\rm (ch)}_v=\frac{\mu_c}{2\pi^2}\bB~~~{\rm and}~~~\bj^{\rm (ch)}_c=\frac{\mu_v}{2\pi^2}\bB\,,
\label{jjBB}}
which is known as the \textit{chiral magnetic effect}. These chemical potentials are defined as $\mu_c\equiv(\mu_++\mu_-)/2$ and $\mu_v\equiv(\mu_+-\mu_-)/2$, in terms of the individual valley chemical potentials $\mu_\pm$.

These chiral magnetic features stem from the Landau-level dispersion of the Weyl fermions subject to a uniform magnetic field [which are easily obtained by squaring the Hamiltonian \rf{H}, in the presence of a static magnetic field $\bB=B\bz$]:
\eq{\al{
\epsilon_0(k_z)&={\rm sgn}(B)\tau vk_z\,,\\
\epsilon_{n\neq0}(k_z)&={\rm sgn}(n)v\sqrt{2|n|/l_B^2+k_z^2}\,,
}\label{ll}}
where $l_B\equiv1/\sqrt{|B|}$ is the magnetic length, $n$ is the integer-valued Landau-level index, and $k_z$ is measured with respect to a Weyl point. Equations \rf{EB} and \rf{jjBB} stem from the linearly-dispersing spectral structure of the zeroth Landau level, $n=0$. Recall that each Landau level has degeneracy $1/2\pi l_B^2$, per unit area in the $xy$ plane.

\section{Dynamic response}
\label{dr}

\subsection{Dynamic variables}

We now wish to construct an effective theory that describes the coupled charge-valley-magnetic dynamics, in the presence of an electromagnetic field. First, let us expand the nonequilibrium dynamics to linear order,\cite{landauBOOKv5} with respect to a uniform thermodynamic equilibrium state with $\bE=0$, $\bB=\textrm{const}$, and $\bM=\textrm{const}$. At low frequencies and long wavelengths, we thus investigate small spatiotemporal modulations of three dynamic variables:  full electron particle density $\rho_c$, valley density $\rho_v$, and collective spin density $s\bm$ (which is allowed to fluctuate slightly with respect to its uniform equilibrium value, focusing on the soft directional dynamics of the unit vector $\bm$).

\subsection{Free energy}

In order to construct the free-energy density $\mF(\rho_c,\rho_v,\bm)$, we start by evaluating the respective conjugate forces $\{\mu_c,\mu_v,\bH\}$ defined by $F_X\equiv\p_X\mF$, for each variable $X=\{\rho_c,\rho_v,\bm\}$. ($\p_X$ is generally understood as a functional derivative.) For degenerate electrons (in a mean-field treatment), in the absence of magnetism, $\mu_c=\rho_c/N+V$, in terms of the Fermi-level density of states $N$ (including both valleys), and $\mu_v=\rho_v/N$. The itinerant-electron contribution to the effective field $\bH$ is given by $\bH_e=\av{\p_\bm H}=-Js\av{\BS}$, where the local spin density $\av{\BS}$ needs to be evaluated over a quasiequilibrium state of the electronic continuum. For the Hamiltonian \rf{H}, it is directly related to the valley flux, since $\BS=\hat{\tau}_z\p_\bk H/v$, so that $\bH_e=-\eta\bj_v$. According to the chiral magnetic effect, Eqs.~\rf{jjBB}, furthermore,
\eq{
\bj_v=\frac{\rho_c}{2\pi^2N}\bB\to v\av{\BS}\,,
\label{jv}}
in equilibrium. This means, in particular,  that the effective spin density $\av{\BS}=\bj_v/v$ is polarized along $\bB$, depending on the charge density $\rho_c$ (relative to the Weyl points).

Integrating the above conjugate forces $\{\mu_c,\mu_v,\bH_e\}$ over the dynamic variables $\{\rho_c,\rho_v,\bm\}$, relative to our uniform reference state, we construct the corresponding free-energy density:
\eq{
\mF_i(\rho_c,\rho_v,\bm)=\frac{1}{2N}\left(\rho_c^2+\rho_v^2-\frac{\eta\rho_c}{\pi^2}\bm\cdot\bB\right)\,,
\label{F}}
to quadratic order in $\{\rho_c,\rho_v\}$ (which are integrated over first) and linear order in $\bm$. Note that the last term in Eq.~\rf{F}, which stems from the magnetic-field induced polarization of the effective spin $\BS$ [cf. Eq.~\rf{jv}], merely shifts the Zeeman coupling of the order parameter $\bm$ in proportion to the electron density $\rho_c$. A similar effect would be present in, e.g., an ordinary Fermi liquid (getting exchange-enhanced on the approach of the Stoner instability).

The full free energy needs also to be supplemented by the electrostatic (Coulombic) contribution, as a functional of $\rho_c(\br)$, and the magnetic contribution (accounting for the Zeeman and dipolar interactions, magnetic anisotropies, etc.), as a functional of $\bm(\br)$, which are generic and unrelated to the Weyl physics. The full free-energy density would correspondingly become:
\eq{
\mF(\rho_c,\rho_v,\bm)=\mF_i(\rho_c,\rho_v,\bm)+\mF_e(\rho_c)+\mF_m(\bm)\,,
\label{mF}}
where $\mF_e$ is the electrostatic and $\mF_m$ the magnetic pieces.

\subsection{Onsager-reciprocal coupling}

Equipped with the quasiequilibrium transport properties captured by Eqs.~\rf{jH}-\rf{jjBB}, we proceed to construct the full coupled equations of motion for our dynamic variables $\{\rho_c,\rho_v,\bm\}(\br,t)$. In the presence of a uniform static magnetic, $\bB$, and a smooth electric, $\bE$, fields (choosing a gauge with $\bE=-\BN V$), these become:
\seq{\an{
\p_t\rho_c+\BN\cdot\bj_c&=0\,,\label{em1}\\
\left(\frac{1}{\tau_v}+\p_t\right)\rho_v+\BN\cdot\bj_v&=\frac{1}{2\pi^2}\bE\cdot\bB\,,\label{em2}\\
s\left(1+\alpha\bm\times\right)\p_t\bm+\bm\times\bH_m&=\boldsymbol{\tau}\,,\label{em3}
}\label{em}}
along with the constitutive relations:
\seq{\an{
\bj_c&=\sigma(1+\theta\bm\times)\bE_c+\frac{\mu_v}{2\pi^2}\bB\,,\label{c1}\\
\bj_v&=-\sigma(\BN\mu_v+\eta\p_t\bm)+\frac{\rho_c}{2\pi^2N}\bB\,,\label{c2}\\
\boldsymbol{\tau}&=\eta\bm\times\bj_v\label{c3}\,.
}\label{c}}
Here, $\bE_c\equiv-\BN\mu_c$, $\sigma$ is the total conductivity,\cite{Note2} $\theta=\eta/2\pi^2\sigma$ the anomalous Hall angle (neglecting the ordinary Hall effect), $\tau_v$ is the valley relaxation time [due to disorder, phonon, or magnon-mediated intervalley scattering, in conjunction with the $\propto\Delta$ term in the Hamiltonian \rf{H}], and $\bH_m\equiv\p_\bm\mF_m$ is the purely magnetic contribution to the effective field $\bH$.

The last two terms $\propto\bB$ in Eqs.~\rf{c1} and \rf{c2} follow from the chiral magnetic effect, Eq.~\rf{jjBB}. The two terms $\propto\eta$ in Eqs.~\rf{c2} and \rf{c3} stem from the valley-dependent vector potential shift \rf{A}:\cite{Note3} The term $\propto\eta$ in Eq.~\rf{c2} originates in the valley-motive force $-{\rm Tr}(\hat{\tau}_z\p_t\bA)/2$.\cite{Note4} The spin torque $\boldsymbol{\tau}$ is obtained as
\eq{
\boldsymbol{\tau}\to-\bm\times\av{\p_\bm H}=Js\bm\times\av{\BS}=\eta\bm\times\bj_v\,,
}
evaluated over a general drift-diffusive state of the electronic system. The three pieces of the valley flux $\bj_v$ in Eq.~\rf{c2}, which contribute to the torque \rf{c3}, enter respectively as the nonequilibrium torque induced by the valley diffusion $\propto\sigma$, Gilbert damping enhancement $\propto\sigma\eta^2$ (which is henceforth absorbed into $\alpha$), and an equilibrium exchange torque $\propto\rho_c$ [corresponding to the free energy \rf{F}]. Inserting Eqs.~\rf{c} into \rf{em}, we establish the coupled charge-valley-magnetic dynamics as:\cite{Note5}
\seq{\an{
\p_t\rho_c+\sigma\BN\cdot\bE_c&=\frac{\bE_v\cdot\bB}{2\pi^2}\,,\label{eom1}\\
\left(\frac{1}{\tau_v}+\p_t\right)\rho_v+\sigma\BN\cdot(\bE_v+\BE)&=\frac{\bE_c\cdot\bB}{2\pi^2}\,,\label{eom2}\\
s\left(1+\alpha\bm\times\right)\p_t\bm+\bm\times(\bH+\bh)&=0\,,\label{eom3}
}\label{eom}}
where $\bE_v\equiv-\BN\mu_v$ and
\seq{\an{
\BE=-\eta(1+\beta\bm\times)\p_t\bm\,,\label{eh1}\\
\bh=-\eta\sigma(1+\beta\bm\times)\bE_v\label{eh2}\,.
}\label{eh}}
Equations~\rf{eom} describe a natural quasistatic relaxation of the perturbed dynamic variables $\{\rho_c,\rho_v,\bm\}$ towards the ground state minimizing the free energy \rf{mF}. The nonelectrical restoring forces, which parametrize the deviation from equilibrium, are explicitly obtained from Eq.~\rf{mF} as:
\eq{
\mu_v=\frac{\rho_v}{N}~~~{\rm and}~~~\bH=\bH_m-\frac{\eta\rho_c}{2\pi^2N}\bB\,.
\label{mmH}}
$\bH_m(\br,t)$ includes the full self-consistent magnetostatic field, exchange stiffness (in the case of an inhomogeneous $\bm$), and any pertinent anisotropies. In the following, we will suppose that the electron liquid is essentially incompressible, so $\rho_c$ is fixed, and the $\propto\eta$ contribution to the magnetic field $\bH$ [Eq.~\rf{mmH}] can be absorbed into $\bH_m$ (and combined with the ordinary Zeeman term).

We wrote the coupled system of Eqs.~\rf{eom} in a form to emphasize the naturally emerging Onsager reciprocity.\cite{landauBOOKv5} The two key physics pieces in these equations, apart from the decoupled valley and charge drift-diffusion and magnetic precession, are (i) the chiral ``anomaly" $\propto\bB$, which engenders the reciprocal coupling between the charge and valley sectors of the dynamics [see the right-hand sides of Eqs.~\rf{eom1} and \rf{eom2}] and (ii) the fictitious gauge potential \rf{A}, $\propto\eta$, which establishes a reciprocal coupling \rf{eh} between the magnetic and valley sectors. These two effects make the motion of the three dynamical variables interdependent.

In addition to the terms present in Eqs.~\rf{em}-\rf{c}, we have phenomenologically supplemented the \textit{field-like} torque $\propto\bm\times\bE_v$ with the \textit{damping-like} torque $\propto\bm\times\bm\times\bE_v$ in Eq.~\rf{eom3}, along with an Onsager-dictated modification of the reciprocal valley-motive force $\BE$ in Eq.~\rf{eom2}, both parametrized by the same dimensionless coefficient $\beta$ [see Eqs.~\rf{eh}]. This is closely analogous to the construction of the field- and damping-like torques induced by the electrical current, along with the reciprocal spin-motive forces, due to the spin Hall effect\cite{tserkovPRB14} (hence the borrowed terminology). We expect such $\beta$ corrections to microscopically relate to the intervalley mixing ($\propto\tau_v^{-1}$) and the associated dissipation. Any local linear torque produced directly by the particle flux $\bj_c$ is ruled out by the inversion symmetry.

To summarize, the theory is constructed by evaluating the \textit{reactive} terms in Eqs.~\rf{em}-\rf{c} explicitly, based on the microscopic model \rf{H}. These enter through a combination of the anomalous quantum Hall effect [$\propto\eta$ in Eq.~\rf{c1}], the chiral anomaly and chiral magnetic effect (all terms $\propto\bB$), and the valley-current torque [$\propto\eta$ in Eq.~\rf{c3}] along with the reciprocal valley-motive force [$\propto\eta$ in Eq.~\rf{c2}]. The \textit{dissipative} terms, parametrized by $\sigma$, $\tau_v^{-1}$, $\alpha$, and $\beta$, are added phenomenologically, on the general grounds of nonequilibrium thermodynamics,\cite{landauBOOKv5} constrained by structural symmetries and Onsager reciprocity.

\section{Examples}
\label{ex}

We will now look at the consequences of the equations of motion derived in the previous section in several illustrative scenarios.

\subsection{Current-induced torques on magnetic textures}

Applying a uniform DC current $\bj_c$ induces the electric field
\eq{
\bE_c\approx\sigma^{-1}\left(1-\theta\bm\times\right)\bj_c\,,
\label{E}}
to the first order in $\theta$ and zeroth order in $\bB$, according to Eq.~\rf{c1}. The generated valley polarization following from Eq.~\rf{em2} is thus
\eq{
\rho_v=\tau_v\frac{\bE_c\cdot\bB}{2\pi^2}\,.
\label{rv}}
Substituting it back into Eq.~\rf{em1}, as $\mu_v=\rho_v/N$, would yield the well-known\cite{andreevPRL18}  positive longitudinal magnetoconductance $\propto\tau_vB^2/N$.

In the presence of a smooth  magnetic texture $\bm(\br)$, the Hall field component \rf{E} results in a modulation of $\rho_v$ [Eq.~\rf{rv}, in turn leading to an effective field $\bh$ [Eq.~\rf{eh2}] as:\cite{Note6}
\eq{
\bh=\eta\sigma(1+\beta\bm\times)\BN\mu_v=\frac{\eta\theta\tau_v}{2\pi^2N}(1+\beta\bm\times)\BN(\bm\cdot\bB\times\bj_c)\,.
\label{h}}
The resultant torque is
\eq{
\boldsymbol{\tau}=\bh\times\bm\propto\bm\times(1+\beta\bm\times)\BN m_x\,,
\label{t}}
where the $x$ axis is oriented along $\bB\times\bj_c$. (We can suppose, for example, that $\bj_c\propto\by$ and $\bB\propto\bz$.) The first term is similar to a conventional Dzyaloshinski-Moriya torque that arises whenever a reflection symmetry is broken along the $x$ axis (as happens, for example, on the surface of a magnetized topological insulator \cite{tserkovPRL12,*tserkovPRB15}). A propensity to form magnetic textures, such as skyrmions, is one consequence of such torques. The second, $\propto\beta$ term in Eq.~\rf{t} describes a Landau-Lifshitz-type damping torque towards the effective field $\propto\BN m_x$. This is reminiscent of the ``chiral damping" discussed in Ref.~\onlinecite{akosaPRB16}.

Similarly to the current-induced torque in ordinary ferromagnets, \cite{bazaliyPRB98,*fernandezPRB04,*tserkovPRB06md} the torque \rf{t} can destabilize a uniform magnetic state. In order to see that, let us look for spin-wave solutions of the Landau-Lifshitz equation \rf{eom3}:
\eq{
s\p_t\bm+\bm\times\bH=\boldsymbol{\tau}_D\,,
}
where the total nonconservative torque is
\eq{
\boldsymbol{\tau}_D=\alpha s\left[\p_t\bm+\nu(1+\beta\bm\times)\BN m_x\right]\times\bm\,,
\label{td}}
where $\nu\propto Bj_c$ absorbed all the relevant torque factors, according to Eq.~\rf{h}. Let the undamped magnetic dynamics have circular precession with dispersion $\omega=b+ak^2$, when expanded with respect to a uniform state $\bm_0\equiv\bz$, and consider a $\nu$-induced instability in a typical situation of $\alpha\ll1$. We can establish when the torque \rf{td} signals an instability by evaluating its work averaged over a cycle of precession $T$:
\eq{\al{
W&=\frac{1}{T}\oint_0^T dt\boldsymbol{\tau}_D\cdot\bm\times\p_t\bm\\
&=\left(\nu\omega\frac{k_x-\beta k_y}{2}-\alpha s\omega^2\right)(\delta m)^2\,,
}}
where $\delta m$ is the spin-wave amplitude. The precession should become unstable, when the right-hand side above becomes positive, resulting in the condition:
\eq{
\nu\frac{k_x-\beta k_y}{2}=\alpha s\omega=\alpha s(b+ak^2)\,.
}
When $\beta\ll1$, the leading instability sets in for $\bk=k\bx$, when $\nu k=2\alpha s(b+ak^2)$. This condition is reached when the two sides of the equation complete the square, which happens for
\eq{
|\nu|=4\alpha s\sqrt{ab}\,,
}
triggering a spin-wave bifurcation at wave number $k={\rm sgn}(\nu)\sqrt{b/a}$. When $\beta\gg1$, the leading instability sets in for $\bk=k\by$ [at $k=-{\rm sgn}(\nu\beta)\sqrt{b/a}$], with a lower threshold of
\eq{
|\nu|=4(\alpha/\beta)s\sqrt{ab}\,.
}

We recall that the $x$ axis is oriented along $\bB\times\bj_c$, which is taken to be normal to the equilibrium state $\bm_0$. When $\beta\to0$, the instability thus affects primarily the waves propagating perpendicular to $\bj_c$ (in contrast to the conventional torques,\cite{bazaliyPRB98} where $\bk\parallel\bj_c$), while when the $\beta$ torque dominates|parallel to $\bj_c$. The valley-mediated torques vanish altogether when $\bB\parallel\bj_c$, which can thus easily be distinguished from the conventional torques, where the relative orientation of $\bB$ and $\bj_c$ is inconsequential. We, furthermore, note that while the larger magnitude of the magnetic field increases $\nu\propto B$ towards the instability here (even when $b\propto B$), the trend is typically opposite for the conventional torques.

\subsection{Magnetic switching}

According to Eq.~\rf{eom3}, the damping-like torque $\propto\beta$ [see Eq.~\rf{eh2}] induces a tendency of the magnetic order $\bm$ to tilt away from $\bE_v$ (supposing $\eta>0$). A large enough $\bE_v$ would thus realize a standard switching for $\bm$, with the final state either pointing antiparallel to $\bE_v$ or undergoing a steady-state precessional dynamics, depending on the magnetic anisotropies.\cite{ralphJMMM08}

One can envision different strategies for inducing $\bE_v$ electrically, based on Eq.~\rf{eom2}. Perhaps most straightforward is to start with the chiral anomaly \rf{rv}, where $\bE_c\approx\bj_c/\sigma$ is controlled by the applied current $\bj_c$ (neglecting the Hall effect here, for simplicity). Near an interface, $\tau_v$ can be strongly reduced by the valley-mixing interfacial scattering, which would induce a drop in the valley polarization of
\eq{
\Delta\rho_v\approx\frac{\tau_v}{2\pi^2\sigma}\bj_c\cdot\bB\,,
}
where $\tau_v$ is the bulk value of the valley-relaxation time. The resultant $\bE_v=-\BN\rho_v/N$, integrated normal to the interface, is
\eq{
\int dr\bE_v\approx\frac{\tau_v}{2\pi^2\sigma N}(\bj_c\cdot\bB)\bn\,,
\label{int}}
where $\bn$ is the outward interface normal (so that $\bj_c\perp\bn$).

Solving the valley-diffusion equation \rf{eom2}, with the boundary condition of $\rho_v\to0$, we see that $\rho_v=\rho_v^{(0)}(1-e^{-r/\lambda_v})$, where $r$ is the distance to the interface and $\rho_v^{(0)}\propto\tau_v\bj_c\cdot\bB$ is the bulk value of the valley polarization (at $r\gg\lambda_v$). The valley-diffusion length (with the diffusion coefficient obeying the Einstein relation, $D=\sigma/N$)
\eq{
\lambda_v=\sqrt{D\tau_v}
\label{lv}}
governs the distance over which the integrated valley field \rf{int} is spread. Having similarly integrated the torque over this distance, we thus find the net damping-like torque of
\eq{
\boldsymbol{\tau}=\eta\sigma\beta\bm\times\bm\times\int dr\bE_v\approx\frac{\eta\beta\tau_v}{2\pi^2N}(\bj_c\cdot\bB)\bm\times(\bm\times\bn)\,,
\label{tn}}
per unit area of the interface. This torque, when large enough. switches $\bm$ towards $\bn$ or $-\bn$, depending on the sign of $\bj_c\cdot\bB$ (with $\bj_c\perp\bn$).

Alternatively, $\bE_v\propto\BN(\tau_v\bj_c\cdot\bB/\sigma)$ can be induced in the bulk by a deliberate modulation of the valley-mixing disorder, affecting $\tau_v(\br)$. Modulating $\sigma$, $\bj_c$, or $\bB$ would result in a similar effect.

\subsection{Magnetic pumping}

According to the valley-motive force $\BE$ [Eq.~\rf{eh1}], the most direct consequence of the magnetic dynamics on our electronic system is valley pumping.\cite{Note7} For a steady periodic precession, 
\eq{
\av{\BE}=-\eta\beta\av{\bm\times\p_t\bm}\to\eta\beta\omega\bz\sin^2\theta\,,
\label{BE}}
where, in the last step, we have specialized to a (left-hand) circular precession around the $z$ axis, with frequency $\omega$ and cone angle $\theta$. This would arise, e.g., as the Larmor precession in response to the magnetic energy $\mF_m=-s\omega\bz\cdot\bm$.

In general, the dynamic valley-motive force $\BE$, which drives the valley current according to Eq.~\rf{eom2}, generates the valley-polarization response described by the valley-diffusion equation (in the absence of the chiral-anomaly term $\propto\bE_c\cdot\bB$):
\eq{
\left(\frac{1}{\tau_v}+\p_t-D\nabla^2\right)\rho_v=-\sigma\BN\cdot\BE\,.
}
Inverting this differential equation, within the bulk, we find
\eq{
\rho_v=-\sigma\int d^3r'\int_{-\infty}^tdt'G(\br-\br',t-t')\BN\cdot\BE(\br',t')\,,
\label{rhov}}
with the Green function describing decaying diffusion:
\eq{
G(\br,t)=\frac{e^{-r^2/4Dt-t/\tau_v}}{(4\pi Dt)^{3/2}}\,.
}

This induced inhomogeneous valley density then produces the source term on the right-hand side of the particle-density continuity equation \rf{eom1}. Supposing the magnetic dynamics are slower than the electrical diffusion/RC response, as is typically the case, we obtain the Gauss law
\eq{
\BN\cdot\bE_c=\frac{\bB\cdot\bs}{2\pi^2N}\,,
\label{GL}}
for the electrochemical field $\bE_c$, where
\eq{
\bs=\int d^3r'\int_{-\infty}^tdt'G(\br-\br',t-t')\BN(\BN\cdot\BE)(\br',t')\,.
}
$\bE_c$ thus follows a Coulomb law with the source obtained from the valley response \rf{rhov}.

In order to find the corresponding voltage generated along a finite-size sample boundary, one needs to revise the above Green function according to the appropriate boundary condition for the valley diffusion (which may account, in particular, for the enhanced valley relaxation at the boundary). For a typical metal, where $\bE_c\approx\bE$ (at long wavelengths), we can conclude, by comparing Eq.~\rf{GL} to the electrostatic Gauss law, $\BN\cdot\bE=4\pi\rho$, that the magnetic dynamics induce the electron charge density of
\eq{
\rho=\frac{\bB\cdot\bs}{8\pi^3N}\,.
}

It is worthwhile reminding that $\mu_c$, which parametrizes $\bE_c\equiv-\BN\mu_c$, stands for the full electrochemical potential built up in the material (here, in response to the magnetic dynamics). In particular, the corresponding voltage developed between different region in the device can be directly measured by making electrical contacts. The characteristic nonlocal character of the electrical response, as reflected in Eqs.~\rf{rhov} and \rf{GL}, should thus be able to provide unambiguous experimental signatures of our phenomenology. When averaged over time, for a cyclic dynamics, only the $\beta$ component of pumping, Eq.~\rf{BE}, remains.

\section{Weyl antiferromagnets}
\label{af}

While we focused our discussion on the ferromagnetic case of the time-reversal symmetry breaking, it is also interesting (and experimentally relevant) to ask about the antiferromagnetic case. Let us consider a bipartite collinear magnet, with two opposite-spin sublattices $A$ and $B$ related by a space-group symmetry $\mM$ (of which we can more generally think as the full subgroup connecting the two sublattices). The N{\'e}el order parameter $\bL=\bM_A-\bM_B$ flips under time reversal and transforms as a magnetization under rotations that do not interchange the sublattices. The sublattice interchange introduces an additional sign change in $\bL$.

If $\mM=\mP$, the parity operator, the magnetism would not break the combined $\mP\mT$ symmetry, thus ruling out the Weyl fermions. We see that also at the level of the Hamiltonian \rf{H}, where the order parameter $\bL$ would not be allowed to replace $\bM$, according to the $\mP$ symmetry. If the $A$ and $B$ sublattices are not interchanged, so that $\bL$ stays invariant under $\mP$, on the other hand, the $\mP\mT$ symmetry gets broken by the antiferromagnetism. The Hamiltonian \rf{H} may now allowed, with the replacement $\bM\to\bL$, as long as it is consistent with $\mM$ under all space-group symmetries.

In general, we need to check how $\mM$ is represented in our four-dimensional basis and whether it constitutes a symmetry of the Hamiltonian \rf{H}. If it does, our phenomenological construction can proceed as in the ferromagnetic case, resulting in the free energy \rf{F} and Onsager-reciprocal equations of motion \rf{eom1} and \rf{eom2}, along with the spin-valley constitutive relations \rf{eh}, with $\bm\to\bl\equiv\bL/L$. The main difference is that the spin torque $\boldsymbol{\tau}=\bh^*\times\bl$ will now enter in the equation of motion for the net antiferromagnetic spin density $d\bS/dt+\dots=\boldsymbol{\tau}$, in lieu of Eq.~\rf{eom3}, where the dots stand for the standard nonlinear $\sigma$-model terms. This must be complemented with the conventional equation of motion for $\bl$, which is dynamically conjugate to $\bS$ (the latter being the generator of rotations in spin space).\cite{tserkovPRB17,*baltzRMP18}

The effective field,
\eq{
\bh^*=\bh-\frac{\eta\rho_c}{2\pi^2N}\bB\,,
}
here includes both the nonequilibrium contribution \rf{eh2} (with $\bm\to\bl$) and the equilibrium one, Eq.~\rf{mmH}, stemming from the free energy \rf{F}. Unlike the ferromagnetic case, where the field $\propto\rho_c\bB$ was qualitatively inconsequential, it can lead to potentially intriguing features in antiferromagnets, where it couples directly to the N{\'e}el order. One interesting possibility concerns the switching of the N{\'e}el order by 180$^\circ$ with $\bB$. Another one is the emergence of the elusive skyrmion crystal in the ground state of an antiferromagnetic material. A linear coupling to the N{\'e}el order is the key missing piece to that end, in conjunction with the Dzyaloshinski-Moriya interaction, which is generically induced by the structural inversion-symmetry breaking.\cite{zarzuelaPRB19}

\section{Summary and discussion}
\label{sd}

Magnetic Weyl semimetals, where magnetism lowers crystalline symmetries and facilitates the formation of topological Weyl points, offer a fertile ground for investigating magnetic dynamics coupled to electrical currents. The key aspect distinguishing such materials from ordinary magnetic metals and semiconductors is the emergence of an approximately conserved valley degree of freedom, which is dictated by the necessary doubling of the Weyl cones.

Here, we have constructed a semi-microscopic phenomenology of the ensuing coupled magneto-valley-electric dynamics. Our concrete microscopic prototype is based on a minimal four-band model, obtained by supplementing the Dirac Hamiltonian with a gap-closing magnetic term. The low-energy dynamics of the valley density, which characterizes an out-of-equilibrium asymmetry in the valley populations, couples to the charge transport as well as magnetic dynamics. The former is rooted in the structure of the chiral anomaly, while the latter stems from the fictitious valley-dependent gauge potential that is engendered by the magnetic order parameter.

Guided by the Onsager reciprocity, the magnetic-valley coupling is further extended to include the leading-order dissipative corrections, which we construct in terms of the valley-motive force, on the one hand, and valley-induced torques, on the other, in the respective equations of motion for the valley and magnetic degrees of freedom. While we mainly focus on the simplest ferromagnetic case, an extension to antiferromagnets is also suggested, which together provide a logical blueprint for other scenarios of magnetic ordering.

Whereas the valley degree of freedom may not be directly accessible to electromagnetic probes, it offers a versatile handle for mediating dynamic magneto-electric phenomena. In particular, we have shown how current-induced spin-wave bifurcations and magnetic switching can be realized in ferromagnet, through the control of the valley-scattering time. In turn, the valley pumping by magnetic dynamics can induce a detectable electrical signal through the chiral magnetic effect. It would be interesting to investigate analogous dynamic phenomena in antiferromagnet and noncollinearly-ordered materials.

Finally, we point out that the Onsager-reciprocal system of the coupled equations of motion like Eqs.~\rf{eom} may generally be constructed on purely phenomenological symmetry-based grounds, starting with any valley-type degrees of freedom in an electronic structure, in conjunction with magnetic or any other order parameter. This can provide a new perspective on magneto-electric phenomena and spintronic functionality based on nontrivial electronic band structures with pockets that are energetically degenerate but spaced in the reciprocal space.

\begin{acknowledgments}
The author is grateful to Shu Zhang and Ji Zou for fruitful discussions. This work was supported by the U.S. Department of Energy, Office of Basic Energy Sciences under Award No.~DE-SC0012190.
\end{acknowledgments}

\end{document}